\documentclass[aps,prb,twocolumn,superscriptaddress,showpacs]{revtex4-1}

\usepackage{graphicx} 
\usepackage{color}
\usepackage{float}

\begin{document}


\title{Magnetic neutron diffraction study of Ba(Fe$_{1-x}$Co$_{x}$)$_{2}$As$_{2}$ critical exponents through the tricritical doping}


\author{D. M. Pajerowski}
\affiliation{NIST Center for Neutron Research, National Institute of Standards and Technology, Gaithersburg, Maryland 20899, USA}

\author{C. R. Rotundu}
\affiliation{Materials Science Division, Lawrence Berkeley National Laboratory, Berkeley, California 94270, USA}

\author{J. W. Lynn}
\affiliation{NIST Center for Neutron Research, National Institute of Standards and Technology, Gaithersburg, Maryland 20899, USA}

\author{R. J. Birgeneau}
\affiliation{Materials Science Division, Lawrence Berkeley National Laboratory, Berkeley, California 94270, USA}
\affiliation{Department of Physics, University of California, Berkeley, California 94270, USA}


\date{\today}

\begin{abstract}
We present temperature dependent magnetic neutron diffraction measurements on Ba(Fe$_{1-x}$Co$_{x}$)$_{2}$As$_{2}$ for $x$ = 0.039, 0.022, and 0.021 as-grown single crystals.  Our  investigations probe the behavior near the magnetic tricritical point in the ($x$,$T$) plane, $x_{tr}~\approx~$0.022, as well as  systematically exploring the character of the magnetic phase transition across a range of doping values.  All samples show long range antiferromagnetic order that may be described near the transition by simple power laws, with $\beta$~=~0.306$\pm$0.060 for $x$~=~0.039, $\beta$~=~0.208$\pm$0.005 for $x$~=~0.022, and $\beta$~=~0.198$\pm$0.009 for $x$~=~0.021.  For the $x$~=~0.039 sample, the data are reasonably well described by the order parameter exponent $\beta$~=~0.326 expected for a 3D Ising model while the $x$~=~0.022 and $x$~=~0.021 samples are near the $\beta$~=~0.25 value for a tricritical system in the mean-field approximation.  These results are discussed in the context of existing experimental work and theoretical predictions.
\end{abstract}


\maketitle



\section{Introduction}

Superconductivity has been a major research interest of the scientific community ever since the first set of experiments in 1911 suggested electron conductance without resistance, and the classes of materials that show a superconducting state have grown extensively over the years.\cite{Matthias1963}  From a technological standpoint, high temperature (high-$T_{C}$) superconductors are particularly attractive, such that when record breaking superconductivity was reported in 1986 for copper oxide based materials\cite{Bednorz1986} the field experienced an enormous surge in activity.\cite{Dagotto1994}  Recently, in 2008,\cite{Kamihara2008} a new paradigm was discovered when iron pnictides were shown to display superconductivity at temperatures in the 50~K range.\cite{Stewart2011}  Both the copper oxide and the iron pnictide systems are characterized by competition between antiferromagnetism and superconductivity.

\begin{figure}[]
	\includegraphics[width=65mm]{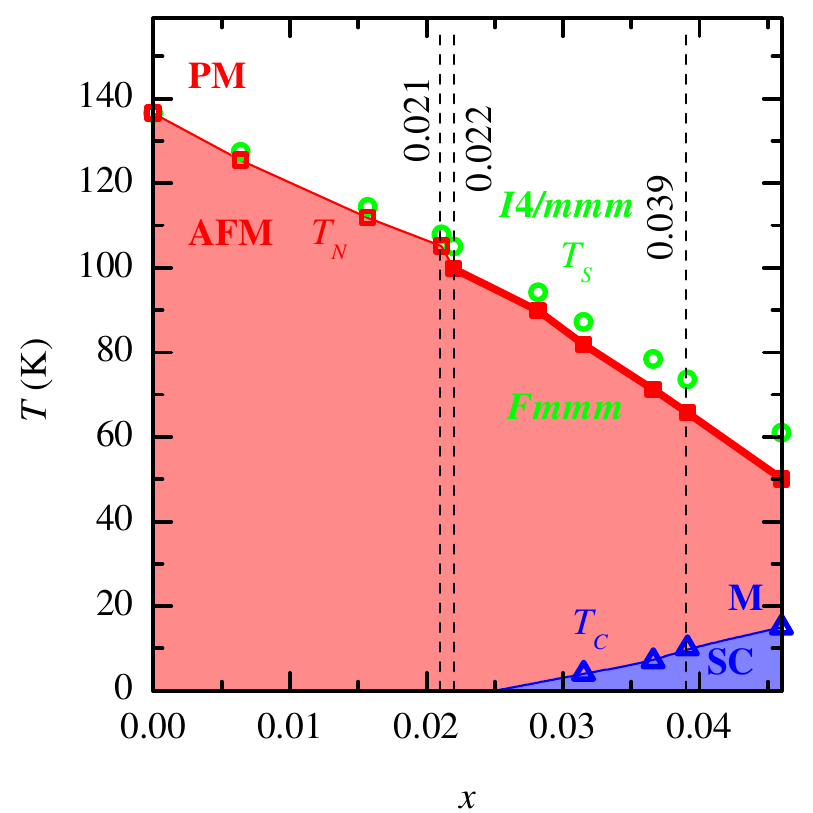}
	\caption{(color online) Ba(Fe$_{1-x}$Co$_{x}$)$_{2}$As$_{2}$ phase diagram in the vicinity of the tricritical point.  The three phases in this region are delineated: the paramagnetic (PM) to antiferromagnetic (AFM) transition at the Neél temperature ($T_{N}$), the tetragonal ($I4/mmm$) to orthorhombic ($Fmmm$) transition at the structural transition temperature ($T_{S}$), and the metallic (M) to superconducting (SC) transition at the critical temperature ($T_{C}$).  For the magnetic transition, open squares correspond to a first order transition and closed squares connected by a solid line correspond to a second order transition.  The vertical dashed lines show the doping levels studied in this work.  The data points for the phase lines are from magnetic susceptibility studies reported in reference \cite{Rotundu2011}.}
	\label{fig:BaFeCoAsFig1}
\end{figure}

With an eye to understanding better the underlying physics in high-$T_{C}$ materials, the iron pnictides quickly became the subject of intense scrutiny.  As a result, a wide variety of iron pnictides, oxypnictides, and chalcogenides have been found to exhibit superconductivity, and it has become common parlance to refer to different structural classes by listing the subscripts of their undoped chemical formulae as in '1111' for compounds isostructural to LaFeAsO, '122' for compounds isostructural to BaFe$_{2}$As$_{2}$, or even more complicated formulae.\cite{Stewart2011}  It is starting to become clear which classes of iron based compounds show technological promise; it is worth mentioning that while cuprates presently retain superiority in terms of high transition temperatures and critical fields, the mechanical properties of some iron based materials which are metallic may better lend themselves to wire production than the more brittle, layered ceramic cuprates.

The interplay of superconductivity and magnetism seen in cuprates is also seen in the iron pnictides, which has motivated concentrated study of the role of the magnetism in this new class of high temperature superconductors.  Interestingly, there is compelling evidence that superconductivity can coexist with long range iron  antiferromagnetic order in the  iron superconductors, while it remains uncertain if copper magnetic order (other than short range) coexists with superconductivity in the cuprates.\cite{Armitage2010}  This coexistence has been shown in some '122' and '1111' iron compounds, where long-range magnetic order and superconductivity arise from bands derived from iron 3d electrons, but is not generic to the system as some formulations do not show coexistence.\cite{Stewart2011}  For the examples without a sharp boundary between superconducting and magnetic phases, the interaction and competition between the phases is seen as a reduction in the ordered moment at the onset of superconductivity.\cite{Pratt2009}  Other non-high-$T_{C}$ magnetic superconductors have shown coexistence, such as borocarbides ($R$Ni$_{2}$B$_{2}$C)\cite{Canfield1998}, the Chevrel phases ($R$Mo$_{6}$S$_{8}$)\cite{Fischer1983}, and ruthenates (RuSr$_{2}$GdCu$_{2}$O$_{8}$)\cite{Lynn2000}, but the most striking analogy is drawn when the same electronic bands participate in both phases such as in UPt$_{3}$\cite{Aeppli1988} and UNi$_{2}$Al$_{3}$\cite{Isaacs1995,Lussier1997}.  Additionally, providing even more evidence for the importance of magnetism in the high-$T_{C}$ iron based superconductors is the observation of a magnetic resonance in the inelastic neutron scattering spectrum, just as previously seen in cuprates and heavy fermion materials.\cite{Lumsden2010}

In the present study, we investigate the cobalt-doped '122' system, Ba(Fe$_{1-x}$Co$_{x}$)$_{2}$As$_{2}$, at low doping.\cite{Sefat2008, Chu2009}  The phase diagram, reproduced using measurements from reference \cite{Rotundu2011} in Fig.~\ref{fig:BaFeCoAsFig1}, shows a typical response to doping.  The parent BaFe$_{2}$As$_{2}$ phase shows two phase transitions when cooling below room temperature, a structural transformation from tetragonal to orthorhombic symmetry (whose temperature is denoted $T_{S}$), and a magnetic transformation from paramagnetic to antiferromagnetic (whose temperature is denoted $T_{N}$).\cite{Rotter2008}  While nearly concurrent in the parent, doping causes a progressive separation of $T_{S}$ and $T_{N}$  with increasing $x$.  Furthermore, upon initial doping, the structural transition is 2nd order and the magnetic transition is 1st order, and, at a doping denoted xtr,  near the region of onset of superconductivity, the structural transition remains 2nd order while the magnetic transition crosses over from 1st to 2nd order.\cite{Kim2011}  This crossover point in the ($x$,$T$) plane, which is near $T$~=~100~K and $x$~=~0.022, is almost certainly a tricritical point.\cite{Rotundu2011}

To explore the nature of the magnetic phase transitions, we have performed temperature dependent neutron diffraction measurements for samples whose Co concentrations, $x$, are in the vicinity of the tricritical value, $x$~=~0.022.  Due to the antiferromagnetic nature of iron based superconductors, neutron scattering has proven ideal for studies of the magnetic structures and excitations.\cite{Lynn2009}  In Sec. II, we outline our experimental procedures for sample preparation and spectrometer configuration.  Section III shows our diffraction data with model fits, while Sec. IV discusses the results of the fits in detail.  Finally, in Sec. V we give our final conclusions and summarize the results.

\section{Experimental Procedure}

\subsection{Synthesis}
Single crystals of Ba(Fe$_{1-x}$Co$_{x}$)$_{2}$As$_{2}$ with cobalt doping values, $x$, of 0.039, 0.022, and 0.021 were grown using a self-flux method, with details available in a previous report.\cite{Rotundu2011}  Samples $x$~=~0.021 of mass 117.2~mg ($T_{N}$~$\approx$~105~K) and $x$~=~0.039 of mass 81.7~mg ($T_{N}$~$\approx$~66~K; $T_{C}$~$\approx$~11~K) are the samples previously used in mapping the phase diagram for cobalt doping \cite{Rotundu2011}, and $x$~=~0.022 of mass 114.6~mg ($T_{N}$~$\approx$~100~K) is from the same batch of the $x$~=~0.022 material in the same study.

\subsection{Instrumentation}
Neutron diffraction experiments were performed on the BT-7 thermal triple-axis spectrometer at the NIST Center for Neutron Research,\cite{Lynn2012} with a collimation of open-50'-sample-50'-120'.  The  $x$~=~0.022 sample was also measured on the high-resolution SPINS cold-source triple-axis spectrometer with a collimation of open-80'-sample-80'-open.  Both machines use the (0 0 2) reflection of pyrolytic graphite (PG) as a monochromator and analyzer.  PG filters to reduce higher order neutrons were employed on the BT-7 spectrometer using a fixed neutron energy of 14.7~meV ($\lambda$~=~2.36~$\textrm{\AA}$), and SPINS used a fixed neutron energy of 5.0~meV ($\lambda$~=~4.05~$\textrm{\AA}$) with a cold Be filter.  Samples were mounted in the $(H~0~L)_{O}$ scattering plane and placed inside a helium flow cryostat, and temperature control was performed in a calibrated geometry capable of at least 50~mK stability.  The energy resolution on BT-7 in this configuration is approximately 1~meV and the energy resolution on SPINS in this configuration is approximately 0.2~meV.  Resolution corrections to the intensity were performed using the Cooper-Nathans approximation.\cite{Cooper1967}  Tabulated values for scattering lengths\cite{Sears1992} and magnetic form factors were used.\cite{Clementi1974}

\section{Neutron Diffraction}
\begin{figure}[b!]
	\includegraphics[width=87mm]{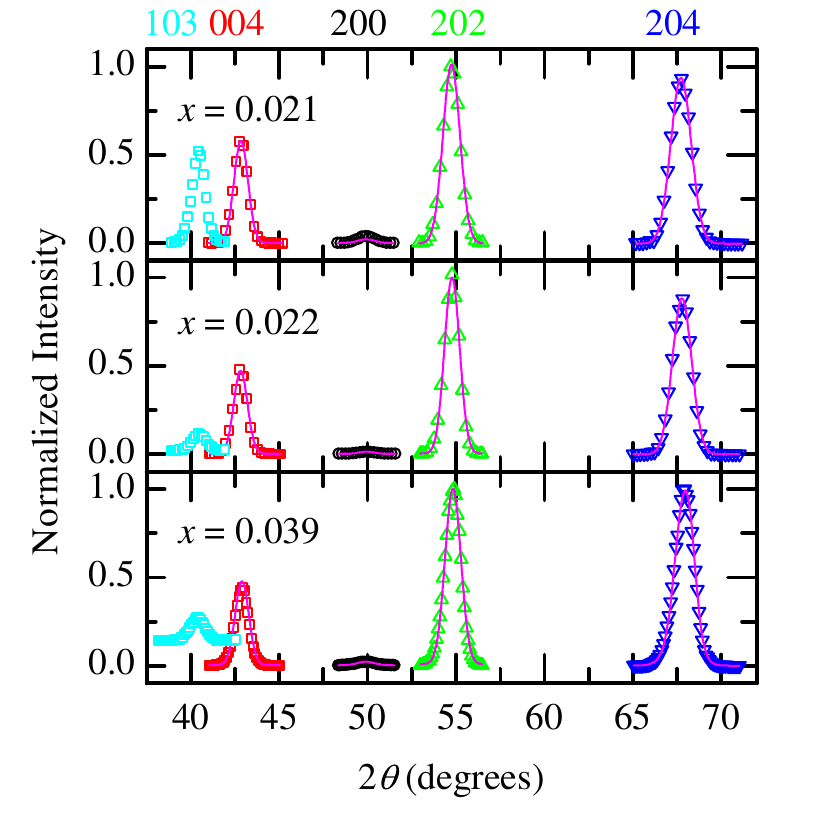}
	\caption{(color online) $\theta-2\theta$  scans in the $(H~0~L)_{O}$ scattering plane for Ba(Fe$_{1-x}$Co$_{x}$)$_{2}$As$_{2}$.  The $(1 0 3)_{O}$ magnetic reflection present at the lowest angle is shown amplified by 50 times for clarity.  The slightly higher background for the $x$~=~0.039 sample is due to a larger detector arm distance used for that measurement.  Uncertainty bars are smaller than the data points, and the lines are results of model fits that are used to extract overall scale factors.}
	\label{fig:BaFeCoAsFig2}
\end{figure}

To begin, we cooled each sample to less than 10~K and performed $\theta-2\theta$ scans of the $(0 0 4)_{O}$, $(2 0 0)_{O}$, $(2 0 2)_{O}$, and $(2 0 4)_{O}$ nuclear reflections as well as the $(1 0 3)_{O}$ magnetic reflection, where  the subscript denotes orthorhombic notation (typical lattice parameters of a~$\approx$~5.62~$\textrm{\AA}$, b~$\approx$~5.57~$\textrm{\AA}$, c~$\approx$~12.94~$\textrm{\AA}$ ).  Typical data are shown in Fig.~\ref{fig:BaFeCoAsFig2}.  The scale-factor was determined from the nuclear peaks using the BaFe2As2 $Fmmm$ (space group No.~69) structure,\cite{Huang2008} with the appropriate substitutions of cobalt for iron in the structure factor calculation.  In these doping ranges, it has been shown that the magnetic structure remains commensurate as in the parent phase.\cite{Kim2010}  In this way, we were able to extract the size of the low-temperature ordered moment to be 0.49$~\mu_{B}\pm$0.01~$\mu_{B}$ ($x$~=~0.021), 0.25$~\mu_{B}\pm$0.01~$\mu_{B}$ ($x$~=~0.022), and 0.31$~\mu_{B}\pm$0.02~$\mu_{B}$ ($x$~=~0.039).

\begin{figure}[b]
	\includegraphics[width=87mm]{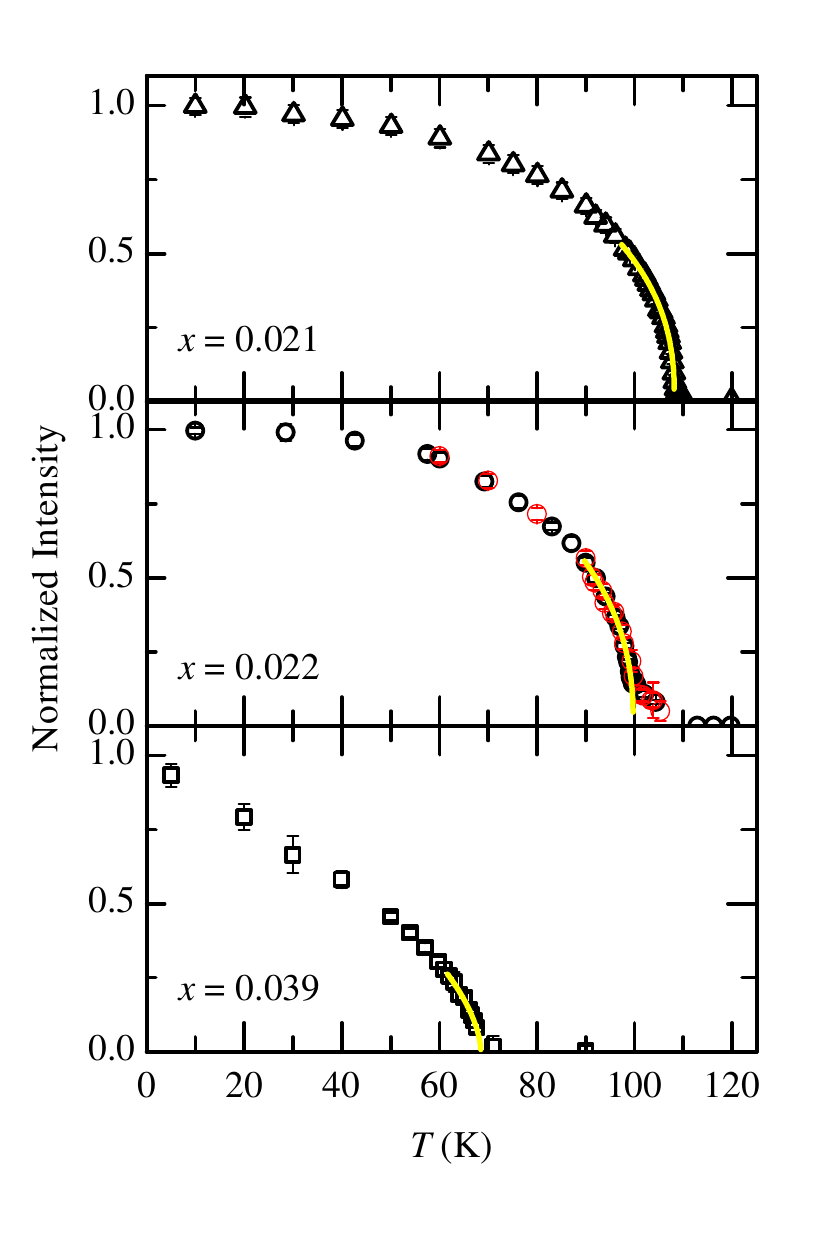}
	\caption{(Color online) Intensity of the magnetic $(1 0 3)_{O}$ peak upon warming.  Experimental data are represented by triangles for $x$~=~0.021, circles for $x$~=~0.022 with both thermal (thick black circles) and cold (thin red circles) neutron diffraction shown, and squares for $x$~=~0.039. The results of model fits are the lines overlaying the data points, details of which are described in the text and in Table I.  The uncertainty bars are derived from counting statistics and represent one standard deviation.}
	\label{fig:BaFeCoAsFig3}
\end{figure}

\begin{table*} []
\caption{Results of fitting to Eq.'s 1-4.\label{table1}}
\begin{ruledtabular}
\begin{tabular}{ c c c c c c c c c c}
\  &	 &	$x$~=~0.021 &	  &	 &	$x$~=~0.022 &	  &	  &	$x$~=~0.039 &	   \\
Eq.  &  $\beta$  &  $T_{N}$ (K)  &  $\sigma$ (K)  &  $\beta$  &  $T_{N}$ (K)  &  $\sigma$ (K)  &  $\beta$  &  $T_{N}$ (K)  &  $\sigma$ (K)   \\
\hline
1   &  0.198$\pm$0.009   &  108.15$\pm$0.09   &  -   &  0.208$\pm$0.005   &  99.73$\pm$0.05   &  -   &  0.306$\pm$0.060   &  68.77$\pm$0.72   &  -  \\
2  &  0.204$\pm$0.006  &  108.32$\pm$0.03  &  0.51$\pm$0.05  &  0.208$\pm$0.006  &  99.73$\pm$0.10  &  0.00$\pm$1.00  &  0.304$\pm$0.087  &  69.03$\pm$1.00  &  1.70$\pm$0.06  \\
3  &  0.296$\pm$0.018  &  108.34$\pm$0.13  &  -  &  0.331$\pm$0.011  &  99.99$\pm$0.07  &  -  &  0.612$\pm$0.178  &  69.64$\pm$1.12  &  -  \\
4  &  0.356$\pm$0.008  &  108.44$\pm$0.05  &  0.44$\pm$0.08  &  0.481$\pm$0.087  &  98.98$\pm$1.05  &  0.76$\pm$0.42  &  0.624$\pm$0.128  &  68.00$\pm$0.97  &  1.29$\pm$0.31  \\
\end{tabular}
\end{ruledtabular}

\end{table*}

In order to understand the critical behavior of the sublattice magnetization as a function of cobalt doping, the temperature dependence of the magnetic $(1 0 3)_{O}$ peak was measured on warming through the magnetic transition for all three samples using $\theta-2\theta$ scans and integrating the intensity.  Near the transition temperature, the intensity, I, of the magnetic diffraction peak may be fit to a simple power law,
\begin{equation}
I~=~A~\left(\frac{T_{N}-T}{T_{N}}\right)^{2\beta}~~~,
\end{equation}
where $T_{N}$ is the Néel temperature, $A$ is a proportionality constant, and $\beta$ is the order parameter critical exponent.  For doped samples like those that we are studying, inhomogeneities in the growth process may give rise to a distribution of Néel temperatures, which if assumed to be Gaussian adds an additional term to eq. 1 for the standard deviation, $\sigma$, such that, for $T<T_{N}$,
\begin{equation}
I~=~A~\int{dt_{N}\frac{1}{\sigma\sqrt{2\pi}}e^{-\frac{1}{2}\left(\frac{t_{N}-T_{N}}{\sigma}\right)^{2}}\left(\frac{t_{N}-T}{t_{N}}\right)^{2\beta}}~~~.
\end{equation}

Finally, renormalization group techniques show that logarithmic corrections lower the effective exponents that one measures when approaching a tricritical point,\cite{Stephen1975} such that the intensity might be modeled as
\begin{equation}
I~=~A~\left(\frac{T_{N}-T}{T_{N}}\right)^{2\beta}~log\left|\frac{T_{N}-T}{T_{N}}\right|^{2\beta}~~~,
\end{equation}
or for the case of distribution of N$\acute{\textrm{e}}$el temperatures, for $T<T_{N}$,
\begin{eqnarray}
I&=&A~\int{dt_{N}}\frac{1}{\sigma\sqrt{2\pi}}e^{-\frac{1}{2}\left(\frac{t_{N}-T_{N}}{\sigma}\right)^{2}} \ldots \nonumber \\ &&\left(\frac{t_{N}-T}{t_{N}}\right)^{2\beta}~log\left|\frac{T_{N}-T}{T_{N}}\right|^{2\beta}~~~.
\end{eqnarray}

\begin{figure} []
	\includegraphics[width=87mm]{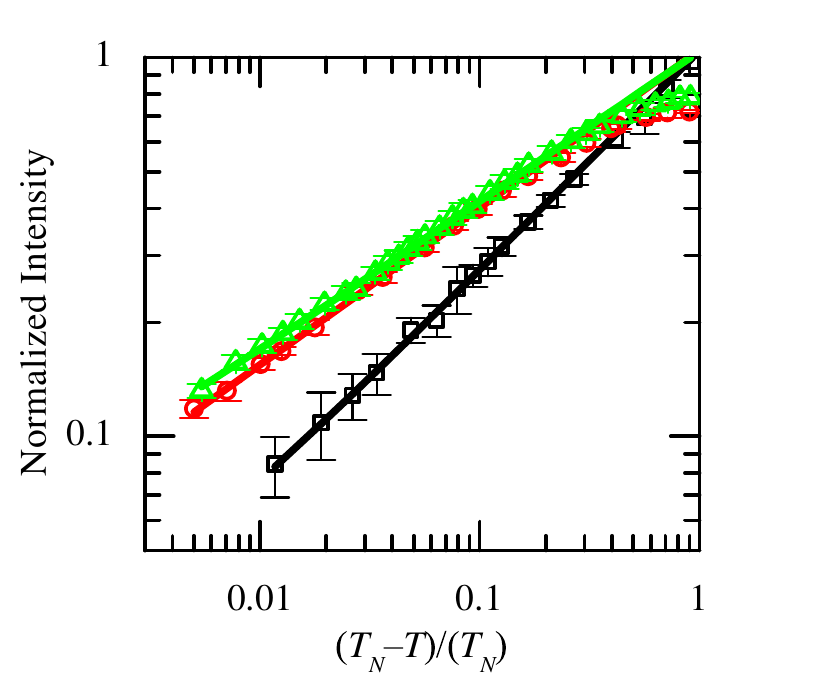}
	\caption{(Color online) Intensity of the magnetic $(1 0 3)_{O}$ peak upon warming on a log-log scale.  The three samples, $x$~=~0.021 (green triangles), $x$~=~0.022 (red circles), and $x$~=~0.039 (black squares) are shown here normalized to their fit functions to illustrate the differences in $/beta$.  The results of model fits to Eq. 1 are lines overlaying the data points, with fit parameters listed in Table I.  Uncertainty bars are derived from counting statistics and represent one standard deviation.}
	\label{fig:BaFeCoAsFig4}
\end{figure}

Results of the measured temperature dependence along with the results of the fits to the data to Eq. 2 are shown in Fig.~\ref{fig:BaFeCoAsFig3} and on a log-log scale in Fig.~\ref{fig:BaFeCoAsFig4}; the best fit parameters for Eq.'s 1-4 are shown in Table I.  In fact, subtle differences between model fits are impossible to discern on the scale of Fig.~\ref{fig:BaFeCoAsFig3} and are obfuscated due to different $T_{N}$ values on plots like Fig.~\ref{fig:BaFeCoAsFig4}, but in the following we elucidate the nuances of each fit.  Fits were performed for reduced temperature within 0.1 of $t_{N}$ to 0.01 of $t_{N}$.  No change in the width of the scattering as a function of temperature was observed.  For the $x$~=~0.022 sample, an abnormal behavior was seen above the preponderant transition temperature that was shown to be elastic by measuring with different energy resolutions of 1 meV and 0.2 meV that showed identical behavior.  This additional scattering was a consequence of the high sensitivity of the transition to cobalt doping; similar behavior was seen in a neutron diffraction study of potassium doped BaFe$_{2}$As$_{2}$.\cite{Rotundu2012}

\begin{figure} [h!]
	\includegraphics[width=87mm]{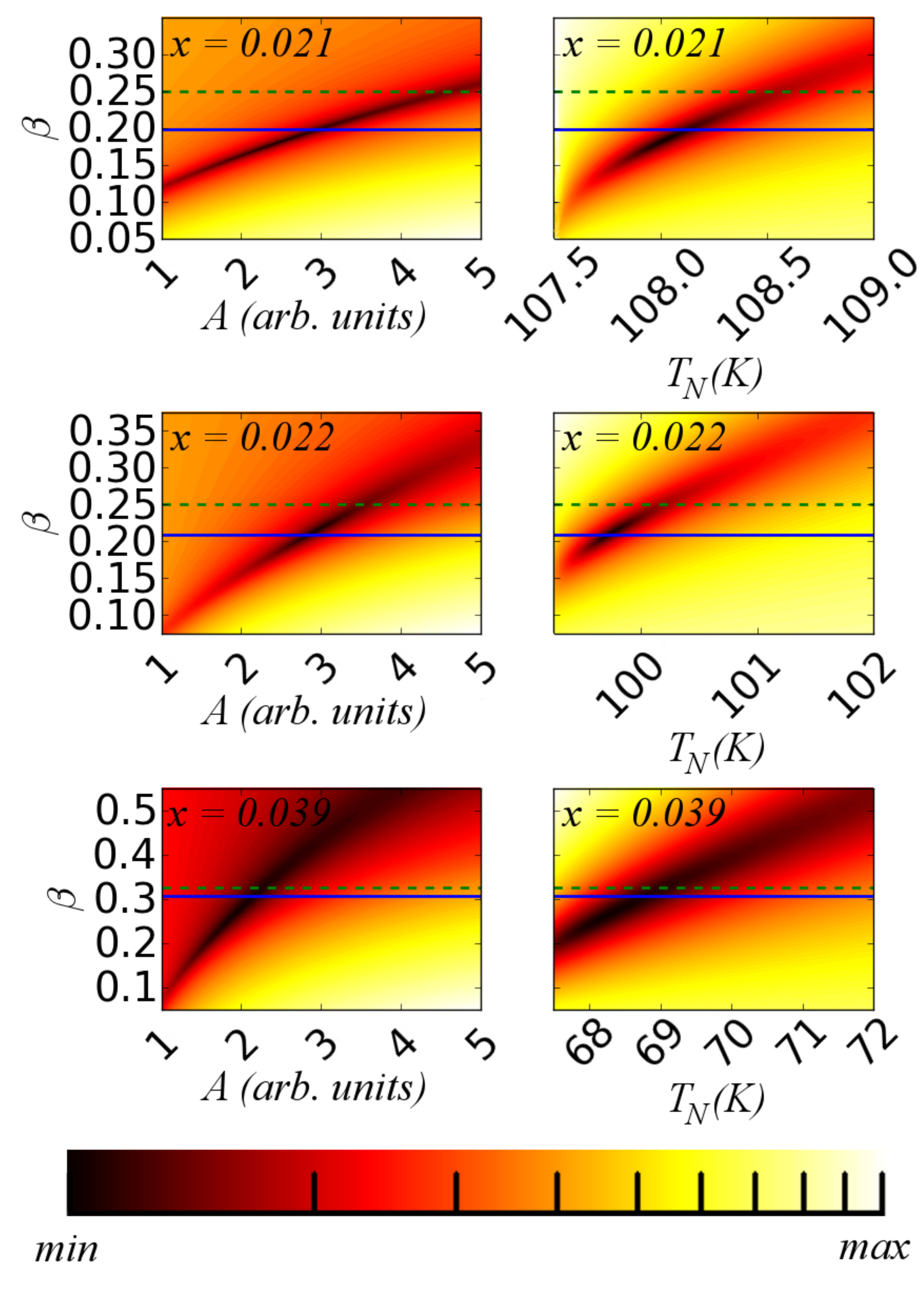}
	\caption{(Color online) (color online) $\chi^{2}$ surfaces from fitting to $I~=~A~\left(\frac{T_{N}-T}{T_{N}}\right)^{2\beta}$.  Using the color map shown at the bottom of this figure, $\chi^{2}$ surfaces on a log scale illustrate the uniqueness of the extracted parameters as well as giving an idea of parameter correlation.  In each subplot, a solid (blue) horizontal line illustrates the best fit $\beta$ value while a dashed (green) horizontal line shows the value for a relevant universality class ($\beta$~=~0.326 of a 3D Ising model for $x$~=~0.039, and $\beta$~=~0.25 of a mean-field tricritical model for $x$~=~0.022 and $x$~=~0.021.)}
	\label{fig:BaFeCoAsFig5}
\end{figure}

Before delving into the physical relevance of the extracted parameters, it is edifying to briefly examine the fits themselves.  The quoted uncertainties are square roots of variances from the least squares algorithm.  To understand how the parameters interact during the fit procedure, the correlation coefficients are a useful, but not necessarily definitive metric,\cite{Spiegel1992} that suggest a strong connection between the proportionality constant, $A$, and the critical exponent, $\beta$, and less so between other parameters.  A clear illustration of possible correlations as well as the goodness of fit are the $\chi^{2}$ surfaces near the solution, which we plot for $\beta$ vs. $T_{N}$ and $\beta$ vs. $A$ for fits to Eq. 1 in Fig.~\ref{fig:BaFeCoAsFig5}.  For a given data set, the four equations have similar $\chi^{2}$ maps, with log corrections (not shown) in eq.'s 3-4 systematically causing a shift to higher $\beta$ values while retaining the basic shape of the minimum.

\section{Discussion}
In the present study, we have measured and analyzed the critical behaviors of the order parameter of cobalt doped barium '122' crystals with a precision approaching the experimental limit dictated by the inherent chemical inhomogeneities characteristic of such doped materials.  We find that the shape of the onset of the magnetic transition to be a function of the amount of doping, in a slightly more complicated way than previously hypothesized for these systems that suggested two distinct critical exponents for '1111' and '122' materials depending upon whether $T_{N}$ and $T_{S}$ were coincident or separated.\cite{Wilson2010}  This general trend of softening the transition with doping is also qualitatively present in a study performed over a large range of doping values.\cite{Fernandes2010}  Quantitatively, for the parent BaFe$_{2}$As$_{2}$, previous neutron diffraction experiments found $\beta$~=~0.103$\pm$0.018, which is less than but near the 2D Ising value of $\beta$~=~0.125.\cite{Wilson2009}  One possibility for such a reduction in the effective exponent is the weakly first order nature of the magnetic transitions for doping values smaller than at the tricritical doping.  Doped Ba(Fe$_{0.953}$Co$_{0.047}$)$_2$As$_2$ ($x$~=~0.047) samples that are within the superconducting range were accurately modeled with $\beta$~=~0.3,[6] which is near the expected value for a 3D Ising model,\cite{Campostrini2002} and we observe a virtually identical value for $x$~=~0.039, Table I.  It is also worth noting that a similar $\beta$-value was seen in nickel doped superconducting samples.\cite{Harriger2009}

In the iron pnictides, the tricritical point is of importance because of its potential role in the onset of superconductivity.  However, there is also a general interest in tricritical points from a fundamental viewpoint, stemming from the inception of the field in the study of $^{3}$He-$^{4}$He mixtures\cite{Griffiths1970} and quickly branching out to magnetic systems.\cite{Birgeneau1974}  Therefore, when additional systems come along that possess a tricritical point, it is exciting to test the validity of the theoretical models.  In cobalt-doped BaFe$_{2}$As$_{2}$, we have studied the tricritical point in the ($x$,$T$) plane, which is expected to behave in a classical manner such that $\beta$~=~0.25.\cite{Landau1935,Huang1987}  As we previously noted, logarithmic corrections to mean-field tricritical exponents have been predicted by theory,\cite{Stephen1975} and such corrections have been applied to similar systems.\cite{Lynn1985}  Nearly identical behavior is seen in the $x$~=~0.021 and $x$~=~0.022 samples that are in the vicinity of the tricritical point.  The best fit exponent for simple power law fits including the evident spread in $T_{N}$ due to chemical inhomogeneity  (Eq.~2) is 0.21$\pm$0.01; this is somewhat less than the mean field tricritical value of 0.25, presumably due to the fact that our data do not probe the true asymptotic critical region.  It is puzzling that the inclusion of a possible logarithmic correction causes the values of beta extracted from the fits to increase dramatically, taking on physically unrealistic values.  For the $x$~=~0.039 sample, where we expect a simple power law to describe the behavior of the order parameter well we find $\beta$=0.30$\pm$0.01, close to, but somewhat less than the 3D Ising value of 0.326.  This is consistent with the results of Wilson and coworkers\cite{Wilson2010} who analyzed similar data in a large number of materials.  It is possible that the reduced effective exponent reflects a residual effect of the nearby tricritical point.

Finally, the moment values that we measure are in the expected range for the doping values measured,\cite{Fernandes2010} although the anomalously low value of the $x$~=~0.022 moment is unexpected.  Previous systematic work has shown appreciable scatter in the moment values, and it is likely that strains in the sample contribute to this distribution.  It is circumstantially evident from the additional features above TN that the $x$~=~0.022 sample may have a larger internal strain.

\section{Conclusions}
In summary, we have measured the critical exponents of Ba(Fe$_{1-x}$Co$_{x}$)$_{2}$As$_{2}$ around and above the tricritical point in the ($x$,$T$) plane, showing behavior consistent with mean-field predictions at the tricritical point.  When above the tricritical point but below optimal doping, we find values of $\beta~\approx$~0.30,  consistent with previous results in a variety of materials but slightly below the expected asymptotic 3D Ising value of 0.326.  As the presence of a tricritical point seems to be a common feature in many iron superconductor systems, it will be interesting to see if other systems show the same behavior.  Finally, there should also be dramatic signatures of the tricritical point in both the heat capacity and the staggered susceptibility.  These properties will be explored in future experiments.

\begin{acknowledgments}
We thank A. Aharony for helpful communications.  We acknowledge the support of the National Institute of Standards and Technology, U.S. Department of Commerce, in providing the neutron research facilities.   DMP acknowledges support from the National Research Council NIST post-doctoral associateship program.  The work at LBL was supported by the Director, Office of Science, Office of Basic Energy Sciences, U.S. Department of Energy, under Contract No. DE-AC02-05CH11231.
\end{acknowledgments}


%


\begin{thebibliography}{22}%
\makeatletter
\providecommand \@ifxundefined [1]{%
 \@ifx{#1\undefined}
}%
\providecommand \@ifnum [1]{%
 \ifnum #1\expandafter \@firstoftwo
 \else \expandafter \@secondoftwo
 \fi
}%
\providecommand \@ifx [1]{%
 \ifx #1\expandafter \@firstoftwo
 \else \expandafter \@secondoftwo
 \fi
}%
\providecommand \natexlab [1]{#1}%
\providecommand \enquote  [1]{``#1''}%
\providecommand \bibnamefont  [1]{#1}%
\providecommand \bibfnamefont [1]{#1}%
\providecommand \citenamefont [1]{#1}%
\providecommand \href@noop [0]{\@secondoftwo}%
\providecommand \href [0]{\begingroup \@sanitize@url \@href}%
\providecommand \@href[1]{\@@startlink{#1}\@@href}%
\providecommand \@@href[1]{\endgroup#1\@@endlink}%
\providecommand \@sanitize@url [0]{\catcode `\\12\catcode `\$12\catcode
  `\&12\catcode `\#12\catcode `\^12\catcode `\_12\catcode `\%12\relax}%
\providecommand \@@startlink[1]{}%
\providecommand \@@endlink[0]{}%
\providecommand \url  [0]{\begingroup\@sanitize@url \@url }%
\providecommand \@url [1]{\endgroup\@href {#1}{\urlprefix }}%
\providecommand \urlprefix  [0]{URL }%
\providecommand \Eprint [0]{\href }%
\providecommand \doibase [0]{http://dx.doi.org/}%
\providecommand \selectlanguage [0]{\@gobble}%
\providecommand \bibinfo  [0]{\@secondoftwo}%
\providecommand \bibfield  [0]{\@secondoftwo}%
\providecommand \translation [1]{[#1]}%
\providecommand \BibitemOpen [0]{}%
\providecommand \bibitemStop [0]{}%
\providecommand \bibitemNoStop [0]{.\EOS\space}%
\providecommand \EOS [0]{\spacefactor3000\relax}%
\providecommand \BibitemShut  [1]{\csname bibitem#1\endcsname}%
\let\auto@bib@innerbib\@empty
\bibitem []{Matthias1963}%
  \BibitemOpen
  \bibfield  {author} 
  {
  \bibinfo {author} {\bibfnamefont {B. T. Matthias}}, 
  \bibinfo {author} {\bibfnamefont {T. H. Geballe}}, 
  \ and
  \bibinfo {author} {\bibfnamefont {V. B. Compton}}, 
  }
  {\bibfield  {journal}
  {
  \bibinfo  {journal} {Rev. Mod. Phys.}\ }
  \textbf {\bibinfo {volume} {35}},\ 
  \bibinfo {pages} {1-22}
  (\bibinfo {year} {1963})
  }\BibitemShut {NoStop}%
\bibitem []{Bednorz1986}%
  \BibitemOpen
  \bibfield  {author} 
  {
  \bibinfo {author} {\bibfnamefont {J. G. Bednorz}} 
  \ and
  \bibinfo {author} {\bibfnamefont {K. A. M$\ddot{\textrm{u}}$ller}}, 
  }
  {\bibfield  {journal}
  {
  \bibinfo  {journal} {Z. Physik B}\ }
  \textbf {\bibinfo {volume} {64}},\ 
  \bibinfo {pages} {189-193}
  (\bibinfo {year} {1986})
  }\BibitemShut {NoStop}%
\bibitem []{Dagotto1994}%
  \BibitemOpen
  \bibfield  {author} 
  {
  \bibinfo {author} {\bibfnamefont {E. Dagotto}}, 
  }
  {\bibfield  {journal}
  {
  \bibinfo  {journal} {Rev. Mod. Phys.}\ }
  \textbf {\bibinfo {volume} {66}},\ 
  \bibinfo {pages} {763-840}
  (\bibinfo {year} {1994})
  }\BibitemShut {NoStop}%
\bibitem []{Kamihara2008}%
  \BibitemOpen
  \bibfield  {author} 
  {
  \bibinfo {author} {\bibfnamefont {Y. Kamihara}}, 
  \bibinfo {author} {\bibfnamefont {T. Watanabe}}, 
  \bibinfo {author} {\bibfnamefont {M. Hirano}}, 
  \ and
  \bibinfo {author} {\bibfnamefont {H. Hosono}}, 
  }
  {\bibfield  {journal}
  {
  \bibinfo  {journal} {J. Am. Chem. Soc.}\ }
  \textbf {\bibinfo {volume} {130}},\ 
  \bibinfo {pages} {3296}
  (\bibinfo {year} {2008})
  }\BibitemShut {NoStop}%
\bibitem []{Stewart2011}%
  \BibitemOpen
  \bibfield  {author} 
  {
  \bibinfo {author} {\bibfnamefont {G. R. Stewart}}, 
  }
  {\bibfield  {journal}
  {
  \bibinfo  {journal} {Rev. Mod. Phys.}\ }
  \textbf {\bibinfo {volume} {83}},\ 
  \bibinfo {pages} {1589-1652}
  (\bibinfo {year} {2011})
  }\BibitemShut {NoStop}%
\bibitem []{Armitage2010}%
  \BibitemOpen
  \bibfield  {author} 
  {
  \bibinfo {author} {\bibfnamefont {N. P. Armitage}}, 
  \bibinfo {author} {\bibfnamefont {R. Fournier}}, 
  \ and
  \bibinfo {author} {\bibfnamefont {R. L. Green}}, 
  }
  {\bibfield  {journal}
  {
  \bibinfo  {journal} {Rev. Mod. Phys.}\ }
  \textbf {\bibinfo {volume} {82}},\ 
  \bibinfo {pages} {2421-2487}
  (\bibinfo {year} {2010})
  }\BibitemShut {NoStop}%
\bibitem []{Pratt2009}%
  \BibitemOpen
  \bibfield  {author} 
  {
  \bibinfo {author} {\bibfnamefont {D. K. Pratt}}, 
  \bibinfo {author} {\bibfnamefont {W. Tian}}, 
  \bibinfo {author} {\bibfnamefont {A. Kreyssig}}, 
  \bibinfo {author} {\bibfnamefont {J. L. Zarestky}}, 
  \bibinfo {author} {\bibfnamefont {S. Nandi}}, 
  \bibinfo {author} {\bibfnamefont {N. Ni}}, 
  \bibinfo {author} {\bibfnamefont {S. L. Bud'ko}}, 
  \bibinfo {author} {\bibfnamefont {P. C. Canfield}}, 
  \bibinfo {author} {\bibfnamefont {A. I. Goldman}}, 
  \ and
  \bibinfo {author} {\bibfnamefont {R. J. McQueeney}}, 
  }
  {\bibfield  {journal}
  {
  \bibinfo  {journal} {Phys. Rev. Lett.}\ }
  \textbf {\bibinfo {volume} {103}},\ 
  \bibinfo {pages} {087001}
  (\bibinfo {year} {2009})
  }\BibitemShut {NoStop}%
\bibitem []{Canfield1998}%
  \BibitemOpen
  \bibfield  {author} 
  {
  \bibinfo {author} {\bibfnamefont {P. C. Canfield}}, 
  \bibinfo {author} {\bibfnamefont {P. L. Gammel}}, 
  \ and
  \bibinfo {author} {\bibfnamefont {D. J. Bishop}}, 
  }
  {\bibfield  {journal}
  {
  \bibinfo  {journal} {Phys. Today}\ }
  \textbf {\bibinfo {volume} {51}},\ 
  \bibinfo {pages} {40}
  (\bibinfo {year} {1998})
  }\BibitemShut {NoStop}%
\bibitem []{Fischer1983}%
  \BibitemOpen
  {\emph {\bibinfo  {title} {Topics in Current Physics}}},\ 
  \bibinfo{editor}{ed.'s $\textrm{\O}$. Fischer and M. B. Maple },
  (\bibinfo  {publisher} {Springer-Verlag},\ 
  \bibinfo {address} {New York},\
  \textbf {\bibinfo {volume} {32, 34}},\ 
  \bibinfo {year} {1983})\BibitemShut {NoStop}%
\bibitem []{Lynn2000}%
  \BibitemOpen
  \bibfield  {author} 
  {
  \bibinfo {author} {\bibfnamefont {J. W. Lynn}}, 
  \bibinfo {author} {\bibfnamefont {B. Keimer}}, 
  \bibinfo {author} {\bibfnamefont {C. Ulrich}}, 
  \bibinfo {author} {\bibfnamefont {C. Bernhard}}, 
  \ and
  \bibinfo {author} {\bibfnamefont {J. L. Tallon}}, 
  }
  {\bibfield  {journal}
  {
  \bibinfo  {journal} {Phys. Rev. B}\ }
  \textbf {\bibinfo {volume} {61}},\ 
  \bibinfo {pages} {R14964}
  (\bibinfo {year} {2000})
  }\BibitemShut {NoStop}%
\bibitem []{Aeppli1988}%
  \BibitemOpen
  \bibfield  {author} 
  {
  \bibinfo {author} {\bibfnamefont {G. Aeppli}}, 
  \bibinfo {author} {\bibfnamefont {E. Bucher}}, 
  \bibinfo {author} {\bibfnamefont {C. Broholm}}, 
  \bibinfo {author} {\bibfnamefont {J. K. Kjems}}, 
  \bibinfo {author} {\bibfnamefont {J. Baumann}}, 
  \ and
  \bibinfo {author} {\bibfnamefont {J. Hufnagl}}, 
  }
  {\bibfield  {journal}
  {
  \bibinfo  {journal} {Phys. Rev. Lett.}\ }
  \textbf {\bibinfo {volume} {60}},\ 
  \bibinfo {pages} {615}
  (\bibinfo {year} {1988})
  }\BibitemShut {NoStop}%
\bibitem []{Isaacs1995}%
  \BibitemOpen
  \bibfield  {author} 
  {
  \bibinfo {author} {\bibfnamefont {E. D. Isaacs}}, 
  \bibinfo {author} {\bibfnamefont {P. Zschack}}, 
  \bibinfo {author} {\bibfnamefont {C. L. Broholm}}, 
  \bibinfo {author} {\bibfnamefont {C. Burns}}, 
  \bibinfo {author} {\bibfnamefont {G. Aeppli}}, 
  \bibinfo {author} {\bibfnamefont {A. P. Ramirez}}, 
  \bibinfo {author} {\bibfnamefont {T. T. M. Palstra}}, 
  \bibinfo {author} {\bibfnamefont {R. W. Erwin}}, 
  \bibinfo {author} {\bibfnamefont {N. Stücheli}}, 
  \ and
  \bibinfo {author} {\bibfnamefont {E. Bucher}}, 
  }
  {\bibfield  {journal}
  {
  \bibinfo  {journal} {Phys. Rev. Lett.}\ }
  \textbf {\bibinfo {volume} {75}},\ 
  \bibinfo {pages} {1178}
  (\bibinfo {year} {1995})
  }\BibitemShut {NoStop}%
\bibitem []{Lussier1997}%
  \BibitemOpen
  \bibfield  {author} 
  {
  \bibinfo {author} {\bibfnamefont {J. G. Lussier}}, 
  \bibinfo {author} {\bibfnamefont {M. Mao}}, 
  \bibinfo {author} {\bibfnamefont {A. Schr$\ddot{\textrm{o}}$der}}, 
  \bibinfo {author} {\bibfnamefont {J. D. Garrett}}, 
  \bibinfo {author} {\bibfnamefont {B. D. Gaulin}}, 
  \bibinfo {author} {\bibfnamefont {S. M. Shapiro}}, 
  \ and
  \bibinfo {author} {\bibfnamefont {W. J. L. Buyers}}, 
  }
  {\bibfield  {journal}
  {
  \bibinfo  {journal} {Phys. Rev. B}\ }
  \textbf {\bibinfo {volume} {56}},\ 
  \bibinfo {pages} {11749}
  (\bibinfo {year} {1997})
  }\BibitemShut {NoStop}%
\bibitem []{Lumsden2010}%
  \BibitemOpen
  \bibfield  {author} 
  {
  \bibinfo {author} {\bibfnamefont {M. D. Lumsden}} 
  \ and
  \bibinfo {author} {\bibfnamefont {A. D. Christianson}}, 
  }
  {\bibfield  {journal}
  {
  \bibinfo  {journal} {J. Phys. Condens. Matter}\ }
  \textbf {\bibinfo {volume} {22}},\ 
  \bibinfo {pages} {203203}
  (\bibinfo {year} {2010})
  }\BibitemShut {NoStop}%
\bibitem []{Sefat2008}%
  \BibitemOpen
  \bibfield  {author} 
  {
  \bibinfo {author} {\bibfnamefont {A. S. Sefat}}, 
  \bibinfo {author} {\bibfnamefont {R. Jin}}, 
  \bibinfo {author} {\bibfnamefont {M. A. McGuire}}, 
  \bibinfo {author} {\bibfnamefont {B. C. Sales}}, 
  \bibinfo {author} {\bibfnamefont {D. J. Singh}}, 
  \ and
  \bibinfo {author} {\bibfnamefont {D. Mandrus}}, 
  }
  {\bibfield  {journal}
  {
  \bibinfo  {journal} {Phys. Rev. Lett.}\ }
  \textbf {\bibinfo {volume} {101}},\ 
  \bibinfo {pages} {117004}
  (\bibinfo {year} {2008})
  }\BibitemShut {NoStop}%
\bibitem []{Chu2009}%
  \BibitemOpen
  \bibfield  {author} 
  {
  \bibinfo {author} {\bibfnamefont {J.-H. Chu}}, 
  \bibinfo {author} {\bibfnamefont {J. G. Analytis}}, 
  \bibinfo {author} {\bibfnamefont {C. Kucharczyk}}, 
  \ and
  \bibinfo {author} {\bibfnamefont {I. R. Fisher}}, 
  }
  {\bibfield  {journal}
  {
  \bibinfo  {journal} {Phys. Rev. B}\ }
  \textbf {\bibinfo {volume} {79}},\ 
  \bibinfo {pages} {014506}
  (\bibinfo {year} {2009})
  }\BibitemShut {NoStop}%
\bibitem []{Rotundu2011}%
  \BibitemOpen
  \bibfield  {author} 
  {
  \bibinfo {author} {\bibfnamefont {C. R. Rotundu}}
  \ and
  \bibinfo {author} {\bibfnamefont {R. J. Birgeneau}}, 
  }
  {\bibfield  {journal}
  {
  \bibinfo  {journal} {Phys. Rev. B}\ }
  \textbf {\bibinfo {volume} {84}},\ 
  \bibinfo {pages} {092501}
  (\bibinfo {year} {2011})
  }\BibitemShut {NoStop}%
\bibitem []{Rotter2008}%
  \BibitemOpen
  \bibfield  {author} 
  {
  \bibinfo {author} {\bibfnamefont {M. Rotter}}, 
  \bibinfo {author} {\bibfnamefont {M. Tegel}}, 
  \bibinfo {author} {\bibfnamefont {D. Johrendt}}, 
  \bibinfo {author} {\bibfnamefont {I. Schellenberg}}, 
  \bibinfo {author} {\bibfnamefont {W. Hermes}}, 
  \ and
  \bibinfo {author} {\bibfnamefont {R. Pottgen}}, 
  }
  {\bibfield  {journal}
  {
  \bibinfo  {journal} {Phys. Rev. B}\ }
  \textbf {\bibinfo {volume} {78}},\ 
  \bibinfo {pages} {020503(R)}
  (\bibinfo {year} {2008})
  }\BibitemShut {NoStop}%
\bibitem []{Kim2011}%
  \BibitemOpen
  \bibfield  {author} 
  {
  \bibinfo {author} {\bibfnamefont {M. G. Kim}}, 
  \bibinfo {author} {\bibfnamefont {R. M. Fernandes}}, 
  \bibinfo {author} {\bibfnamefont {A. Kreyssig}}, 
  \bibinfo {author} {\bibfnamefont {J. W. Kim}}, 
  \bibinfo {author} {\bibfnamefont {A. Thaler}}, 
  \bibinfo {author} {\bibfnamefont {S. L. Bud'ko}}, 
  \bibinfo {author} {\bibfnamefont {P. C. Canfield}}, 
  \bibinfo {author} {\bibfnamefont {R. J. McQueeney}}, 
  \bibinfo {author} {\bibfnamefont {J. Schmalian}}, 
  \ and
  \bibinfo {author} {\bibfnamefont {A. I. Goldman}}, 
  }
  {\bibfield  {journal}
  {
  \bibinfo  {journal} {Phys. Rev. B}\ }
  \textbf {\bibinfo {volume} {83}},\ 
  \bibinfo {pages} {134522}
  (\bibinfo {year} {2011})
  }\BibitemShut {NoStop}%
\bibitem []{Mathur1998}%
  \BibitemOpen
  \bibfield  {author} 
  {
  \bibinfo {author} {\bibfnamefont {N. D. Mathur}}, 
  \bibinfo {author} {\bibfnamefont {F. M. Grosche}}, 
  \bibinfo {author} {\bibfnamefont {S. R. Julian}}, 
  \bibinfo {author} {\bibfnamefont {I. R. Walker}}, 
  \bibinfo {author} {\bibfnamefont {D. M. Freye}}, 
  \bibinfo {author} {\bibfnamefont {R. K. W. Haselwimmer}}, 
  \ and
  \bibinfo {author} {\bibfnamefont {G. G. Lonzarich}}, 
  }
  {\bibfield  {journal}
  {
  \bibinfo  {journal} {Nature}\ }
  \textbf {\bibinfo {volume} {394}},\ 
  \bibinfo {pages} {39-43}
  (\bibinfo {year} {1998})
  }\BibitemShut {NoStop}%
\bibitem []{Sachdev2010}%
  \BibitemOpen
  \bibfield  {author} 
  {
  \bibinfo {author} {\bibfnamefont {S. Sachdev}}, 
  }
  {\bibfield  {journal}
  {
  \bibinfo  {journal} {Phys. Stat. Sol. (b)}\ }
  \textbf {\bibinfo {volume} {247}},\ 
  \bibinfo {pages} {537-543}
  (\bibinfo {year} {2010})
  }\BibitemShut {NoStop}%
\bibitem []{Dai2009}%
  \BibitemOpen
  \bibfield  {author} 
  {
  \bibinfo {author} {\bibfnamefont {J. Dai}}, 
  \bibinfo {author} {\bibfnamefont {Z. Si}}, 
  \bibinfo {author} {\bibfnamefont {J. -X. Zhu}}, 
  \ and
  \bibinfo {author} {\bibfnamefont {E. Abrahams}}, 
  }
  {\bibfield  {journal}
  {
  \bibinfo  {journal} {Proc. Nat. Acad. Sci. USA}\ }
  \textbf {\bibinfo {volume} {106}},\ 
  \bibinfo {pages} {4118-4121}
  (\bibinfo {year} {2009})
  }\BibitemShut {NoStop}%
\bibitem []{Giovannetti2011}%
  \BibitemOpen
  \bibfield  {author} 
  {
  \bibinfo {author} {\bibfnamefont {G. Giovannetti}}, 
  \bibinfo {author} {\bibfnamefont {C. Ortix}}, 
  \bibinfo {author} {\bibfnamefont {M. Marsman}}, 
  \bibinfo {author} {\bibfnamefont {M. Capone}}, 
  \bibinfo {author} {\bibfnamefont {J. vandenBrink}}, 
  \ and
  \bibinfo {author} {\bibfnamefont {J. Lorenzana}}, 
  }
  {\bibfield  {journal}
  {
  \bibinfo  {journal} {Nat. Commun.}\ }
  \textbf {\bibinfo {volume} {2}},\ 
  \bibinfo {pages} {398}
  (\bibinfo {year} {2011})
  }\BibitemShut {NoStop}%
\bibitem []{Lynn2009}%
  \BibitemOpen
  \bibfield  {author} 
  {
  \bibinfo {author} {\bibfnamefont {J. W. Lynn}}
  \ and
  \bibinfo {author} {\bibfnamefont {P. Dai}}, 
  }
  {\bibfield  {journal}
  {
  \bibinfo  {journal} {Physica C}\ }
  \textbf {\bibinfo {volume} {469}},\ 
  \bibinfo {pages} {469}
  (\bibinfo {year} {2009})
  }\BibitemShut {NoStop}%
\bibitem []{Lynn2012}%
  \BibitemOpen
  \bibfield  {author} 
  {
  \bibinfo {author} {\bibfnamefont {J. W. Lynn}}, 
  \bibinfo {author} {\bibfnamefont {Y. Chen}}, 
  \bibinfo {author} {\bibfnamefont {S. Chang}}, 
  \bibinfo {author} {\bibfnamefont {Y. Zhao}}, 
  \bibinfo {author} {\bibfnamefont {S. Chi}}, 
  \bibinfo {author} {\bibfnamefont {W. Ratcliff}}, 
  \bibinfo {author} {\bibfnamefont {B. G. Ueland}}, 
  \ and
  \bibinfo {author} {\bibfnamefont {R. W. Erwin}}, 
  }
  {\bibfield  {journal}
  {
  \bibinfo  {journal} {J. Res. Natl. Inst. Stan.}\ }
  \textbf {\bibinfo {volume} {117}},\ 
  \bibinfo {pages} {61-79}
  (\bibinfo {year} {2012})
  }\BibitemShut {NoStop}%
\bibitem []{Cooper1967}%
  \BibitemOpen
  \bibfield  {author} 
  {
  \bibinfo {author} {\bibfnamefont {M. J. Cooper}}
  \ and
  \bibinfo {author} {\bibfnamefont {R. Nathans}}, 
  }
  {\bibfield  {journal}
  {
  \bibinfo  {journal} {Acta Cryst.}\ }
  \textbf {\bibinfo {volume} {23}},\ 
  \bibinfo {pages} {367-376}
  (\bibinfo {year} {1967})
  }\BibitemShut {NoStop}%
\bibitem []{Sears1992}%
  \BibitemOpen
  \bibfield  {author} 
  {
  \bibinfo {author} {\bibfnamefont {V. F. Sears}}, 
  }
  {\bibfield  {journal}
  {
  \bibinfo  {journal} {Neutron News}\ }
  \textbf {\bibinfo {volume} {3}},\ 
  \bibinfo {pages} {26-27}
  (\bibinfo {year} {1992})
  }\BibitemShut {NoStop}%
\bibitem []{Clementi1974}%
  \BibitemOpen
  \bibfield  {author} 
  {
  \bibinfo {author} {\bibfnamefont {E. Clementi}} 
  \ and
  \bibinfo {author} {\bibfnamefont {C. Roetti}}, 
  }
  {\bibfield  {journal}
  {
  \bibinfo  {journal} {Atom Data Nucl. Data}\ }
  \textbf {\bibinfo {volume} {14}},\ 
  \bibinfo {pages} {177}
  (\bibinfo {year} {1974})
  }\BibitemShut {NoStop}%
\bibitem []{Huang2008}%
  \BibitemOpen
  \bibfield  {author} 
  {
  \bibinfo {author} {\bibfnamefont {Q. Huang}}, 
  \bibinfo {author} {\bibfnamefont {Y. Qiu}}, 
  \bibinfo {author} {\bibfnamefont {W. Bao}}, 
  \bibinfo {author} {\bibfnamefont {M. A. Green}}, 
  \bibinfo {author} {\bibfnamefont {J. W. Lynn}}, 
  \bibinfo {author} {\bibfnamefont {Y. C. Gasparovic}}, 
  \bibinfo {author} {\bibfnamefont {T. Wu}}, 
  \bibinfo {author} {\bibfnamefont {G. Wu}}, 
  \ and
  \bibinfo {author} {\bibfnamefont {X. H. Chen}}, 
  }
  {\bibfield  {journal}
  {
  \bibinfo  {journal} {Phys. Rev. Lett.}\ }
  \textbf {\bibinfo {volume} {101}},\ 
  \bibinfo {pages} {257003}
  (\bibinfo {year} {2008})
  }\BibitemShut {NoStop}%
\bibitem []{Kim2010}%
  \BibitemOpen
  \bibfield  {author} 
  {
  \bibinfo {author} {\bibfnamefont {M. G. Kim}}, 
  \bibinfo {author} {\bibfnamefont {A. Kreyssig}}, 
  \bibinfo {author} {\bibfnamefont {Y. B. Lee}}, 
  \bibinfo {author} {\bibfnamefont {J. W. Kim}}, 
  \bibinfo {author} {\bibfnamefont {D. K. Pratt}}, 
  \bibinfo {author} {\bibfnamefont {A. Thaler}}, 
  \bibinfo {author} {\bibfnamefont {S. L. Bud'ko}}, 
  \bibinfo {author} {\bibfnamefont {P. C. Canfield}}, 
  \bibinfo {author} {\bibfnamefont {B. N. Harmon}}, 
  \bibinfo {author} {\bibfnamefont {R. J. McQueeney}}, 
  \ and
  \bibinfo {author} {\bibfnamefont {A. I. Goldman}}, 
  }
  {\bibfield  {journal}
  {
  \bibinfo  {journal} {Phys. Rev. B}\ }
  \textbf {\bibinfo {volume} {82}},\ 
  \bibinfo {pages} {180412}
  (\bibinfo {year} {2010})
  }\BibitemShut {NoStop}%
\bibitem []{Stephen1975}%
  \BibitemOpen
  \bibfield  {author} 
  {
  \bibinfo {author} {\bibfnamefont {M. J. Stephen}}, 
  \bibinfo {author} {\bibfnamefont {E. Abrahams}}, 
  \ and
  \bibinfo {author} {\bibfnamefont {J. P. Straley}}, 
  }
  {\bibfield  {journal}
  {
  \bibinfo  {journal} {Phys. Rev. B}\ }
  \textbf {\bibinfo {volume} {12}},\ 
  \bibinfo {pages} {256–262}
  (\bibinfo {year} {1975})
  }\BibitemShut {NoStop}%
\bibitem []{Rotundu2012}%
  \BibitemOpen
  \bibfield  {author} 
  {
  \bibinfo {author} {\bibfnamefont {C. R. Rotundu}}, 
  \bibinfo {author} {\bibfnamefont {W. Tian}}, 
  \bibinfo {author} {\bibfnamefont {K. C. Rule}}, 
  \bibinfo {author} {\bibfnamefont {T. R. Forrest}}, 
  \bibinfo {author} {\bibfnamefont {J. Zhao}}, 
  \bibinfo {author} {\bibfnamefont {J. L. Zarestky}}, 
  \ and
  \bibinfo {author} {\bibfnamefont {R. J. Birgeneau}}, 
  }
  {\bibfield  {journal}
  {
  \bibinfo  {journal} {Phys. Rev. B}\ }
  \textbf {\bibinfo {volume} {85}},\ 
  \bibinfo {pages} {144506}
  (\bibinfo {year} {2012})
  }\BibitemShut {NoStop}%
\bibitem []{Spiegel1992}%
  \BibitemOpen
  \bibfield  {author} 
  {
  \bibinfo {author} {\bibfnamefont {M. R. Spiegel}}, 
  }
  {\emph {\bibinfo  {title} {Theory and Problems of Probability and Statistics, 2nd ed.}}},\ 
  (\bibinfo  {publisher} {McGraw-Hill},\ 
  \bibinfo {address} {New York},\
  \bibinfo {year} {1992})\BibitemShut {NoStop}%
\bibitem []{Wilson2010}%
  \BibitemOpen
  \bibfield  {author} 
  {
  \bibinfo {author} {\bibfnamefont {S. D. Wilson}}, 
  \bibinfo {author} {\bibfnamefont {C. R. Rotundu}}, 
  \bibinfo {author} {\bibfnamefont {Z. Yamani}}, 
  \bibinfo {author} {\bibfnamefont {P. Valdivia}}, 
  \bibinfo {author} {\bibfnamefont {B. Freelon}}, 
  \bibinfo {author} {\bibfnamefont {E. Bourret-Courchesne}}, 
  \ and
  \bibinfo {author} {\bibfnamefont {R. J. Birgneneau}}, 
  }
  {\bibfield  {journal}
  {
  \bibinfo  {journal} {Phys. Rev. B}\ }
  \textbf {\bibinfo {volume} {81}},\ 
  \bibinfo {pages} {014501}
  (\bibinfo {year} {2010})
  }\BibitemShut {NoStop}%
\bibitem []{Fernandes2010}%
  \BibitemOpen
  \bibfield  {author} 
  {
  \bibinfo {author} {\bibfnamefont {R. M. Fernandes}}, 
  \bibinfo {author} {\bibfnamefont {D. K. Pratt}}, 
  \bibinfo {author} {\bibfnamefont {W. Tian}}, 
  \bibinfo {author} {\bibfnamefont {J. Zarestky}}, 
  \bibinfo {author} {\bibfnamefont {A. Kreyssig}}, 
  \bibinfo {author} {\bibfnamefont {S. Nandi}}, 
  \bibinfo {author} {\bibfnamefont {M. G. Kim}}, 
  \bibinfo {author} {\bibfnamefont {A. Thaler}}, 
  \bibinfo {author} {\bibfnamefont {N. Ni}}, 
  \bibinfo {author} {\bibfnamefont {P. C. Canfield}}, 
  \bibinfo {author} {\bibfnamefont {R. J. McQueeney}}, 
  \bibinfo {author} {\bibfnamefont {J. Schmalian}}, 
  \ and
  \bibinfo {author} {\bibfnamefont {A. I. Goldman}}, 
  }
  {\bibfield  {journal}
  {
  \bibinfo  {journal} {Phys. Rev. B}\ }
  \textbf {\bibinfo {volume} {81}},\ 
  \bibinfo {pages} {140501(R)}
  (\bibinfo {year} {2010})
  }\BibitemShut {NoStop}%
\bibitem []{Wilson2009}%
  \BibitemOpen
  \bibfield  {author} 
  {
  \bibinfo {author} {\bibfnamefont {S. D. Wilson}}, 
  \bibinfo {author} {\bibfnamefont {Z. Yamani}}, 
  \bibinfo {author} {\bibfnamefont {C. R. Rotundu}}, 
  \bibinfo {author} {\bibfnamefont {B. K. Freelon}}, 
  \bibinfo {author} {\bibfnamefont {E. B. C. }}, 
  \ and
  \bibinfo {author} {\bibfnamefont {R. J. Birgeneau}}, 
  }
  {\bibfield  {journal}
  {
  \bibinfo  {journal} {Phys. Rev. B}\ }
  \textbf {\bibinfo {volume} {79}},\ 
  \bibinfo {pages} {184519}
  (\bibinfo {year} {2009})
  }\BibitemShut {NoStop}%
\bibitem []{Campostrini2002}%
  \BibitemOpen
  \bibfield  {author} 
  {
  \bibinfo {author} {\bibfnamefont {M. Campostrini}}, 
  \bibinfo {author} {\bibfnamefont {A. Pelissetto}}, 
  \bibinfo {author} {\bibfnamefont {P. Rossi}}, 
  \ and
  \bibinfo {author} {\bibfnamefont {E. Vicari}}, 
  }
  {\bibfield  {journal}
  {
  \bibinfo  {journal} {Phys. Rev. E}\ }
  \textbf {\bibinfo {volume} {65}},\ 
  \bibinfo {pages} {066127}
  (\bibinfo {year} {2002})
  }\BibitemShut {NoStop}%
\bibitem []{Harriger2009}%
  \BibitemOpen
  \bibfield  {author} 
  {
  \bibinfo {author} {\bibfnamefont {L. W. Harriger}}, 
  \bibinfo {author} {\bibfnamefont {A. Schneidewind}}, 
  \bibinfo {author} {\bibfnamefont {S. Li}}, 
  \bibinfo {author} {\bibfnamefont {J. Zhao}}, 
  \bibinfo {author} {\bibfnamefont {Z. Li}}, 
  \bibinfo {author} {\bibfnamefont {W. Lu}}, 
  \bibinfo {author} {\bibfnamefont {X. Dong}}, 
  \bibinfo {author} {\bibfnamefont {F. Zhou}}, 
  \bibinfo {author} {\bibfnamefont {Z. Zhao}}, 
  \bibinfo {author} {\bibfnamefont {J. Hu}}, 
  \ and
  \bibinfo {author} {\bibfnamefont {P. Dai}}, 
  }
  {\bibfield  {journal}
  {
  \bibinfo  {journal} {Phys. Rev. Lett.}\ }
  \textbf {\bibinfo {volume} {103}},\ 
  \bibinfo {pages} {087005}
  (\bibinfo {year} {2009})
  }\BibitemShut {NoStop}%
\bibitem []{Griffiths1970}%
  \BibitemOpen
  \bibfield  {author} 
  {
  \bibinfo {author} {\bibfnamefont {R. B. Griffiths}}, 
  }
  {\bibfield  {journal}
  {
  \bibinfo  {journal} {Phys. Rev. Lett.}\ }
  \textbf {\bibinfo {volume} {24}},\ 
  \bibinfo {pages} {715}
  (\bibinfo {year} {1970})
  }\BibitemShut {NoStop}%
\bibitem []{Birgeneau1974}%
  \BibitemOpen
  \bibfield  {author} 
  {
  \bibinfo {author} {\bibfnamefont {R. J. Birgeneau}}, 
  \bibinfo {author} {\bibfnamefont {G. Shirane}}, 
  \bibinfo {author} {\bibfnamefont {M. Blume}}, 
  \ and
  \bibinfo {author} {\bibfnamefont {W. C. Koehler}}, 
  }
  {\bibfield  {journal}
  {
  \bibinfo  {journal} {Phys. Rev. Lett.}\ }
  \textbf {\bibinfo {volume} {33}},\ 
  \bibinfo {pages} {1098}
  (\bibinfo {year} {1974})
  }\BibitemShut {NoStop}%
\bibitem []{Landau1935}%
  \BibitemOpen
  \bibfield  {author} 
  {
  \bibinfo {author} {\bibfnamefont {L. D. Landau}}, 
  }
  {\bibfield  {journal}
  {
  \bibinfo  {journal} {Phys. Z. Sowjetunion}\ }
  \textbf {\bibinfo {volume} {8}},\ 
  \bibinfo {pages} {113}
  (\bibinfo {year} {1935})
  }\BibitemShut {NoStop}%
\bibitem []{Huang1987}%
  \BibitemOpen
  \bibfield  {author} 
  {
  \bibinfo {author} {\bibfnamefont {K. Huang}}, 
  }
  {\emph {\bibinfo  {title} {Statistical Mechanics, 2nd ed.}}},\ 
  (\bibinfo  {publisher} {Wiley},\ 
  \bibinfo {address} {New York},\
  \bibinfo {pages} {432-438},\
  \bibinfo {year} {1987})\BibitemShut {NoStop}%
\bibitem []{Lynn1985}%
  \BibitemOpen
  \bibfield  {author} 
  {
  \bibinfo {author} {\bibfnamefont {J. W. Lynn}}, 
  \bibinfo {author} {\bibfnamefont {J. A. Gotaas}}, 
  \bibinfo {author} {\bibfnamefont {R. N. Shelton}}, 
  \bibinfo {author} {\bibfnamefont {H. E. Horng}}, 
  \ and
  \bibinfo {author} {\bibfnamefont {C. J. Glinka}}, 
  }
  {\bibfield  {journal}
  {
  \bibinfo  {journal} {Phys. Rev. B}\ }
  \textbf {\bibinfo {volume} {31}},\ 
  \bibinfo {pages} {5756}
  (\bibinfo {year} {1985})
  }\BibitemShut {NoStop}%
%
%
%
\end{thebibliography}
\end{document}